# Harmonic oscillator with time-dependent effective-mass and frequency with a possible application to 'chirped tidal' gravitational waves forces affecting interferometric detectors

Yacob Ben-Aryeh

Physics Department, Technion-Israel Institute of Technology, Haifa,32000,Israel

e-mail: phr65yb@physics.technion.ac.il; Fax: 972-4-8295755

**Abstract**

The general theory of time-dependent frequency and time-dependent mass ('effective mass') is described. The general theory for time-dependent harmonic-oscillator is applied in the present research for studying certain quantum effects in the interferometers for detecting gravitational waves. When an astronomical binary system approaches its point of coalescence the gravitational wave intensity and frequency are increasing and this can lead to strong deviations from the simple description of harmonic oscillations for the interferometric masses on which the mirrors are placed. It is shown that under such conditions the harmonic oscillations of these masses can be described by mechanical harmonic-oscillators with time-dependent frequency and effective-mass. In the present theoretical model the effective-mass is decreasing with time describing pumping phenomena in which the oscillator amplitude is increasing with time. The quantization of this system is analyzed by the use of the adiabatic approximation. It is found that the increase of the gravitational wave intensity, within the adiabatic approximation, leads to squeezing phenomena where the quantum noise in one quadrature is increased and in the other quadrature it is decreased.

## 1. Introduction

The problem of harmonic-oscillator with time-dependent mass has been related to a quantum damped oscillator [1-7]. In these studies the mass parameter is given as a general function of time, and the equations of motion include damping mechanism with a driving force [7] or without it [6]. As these systems are non-conservative, quantum theories of non-conservative systems have been developed.

The study of problems involving harmonic oscillators with time-dependent masses [8-9], or with time-dependent frequencies [10-19], or both simultaneously [20-28] has attracted much interest with the various publications. Of course such systems are not closed in the sense that some external influence, which not need be specified, may change the harmonic oscillator parameters, i.e., alter its frequency, amplitude etc. By using the generalized quantum mechanical invariant, first introduced by Lewis [13,14] and Lewis and Reisenfeld [11], the exact quantum states for such oscillators can be found for various special cases. It is shown that squeezing phenomena (i.e., quantum fluctuations in one quadrature which are smaller relative to those of the other quadrature) will inevitably be generated in a harmonic oscillator with time dependent-frequency and mass. Squeezing phenomena have been found in the analysis of harmonic-oscillators with variable mass [8,9] or variable frequency [10-19], or with both [20-28]. Variable frequencies in electromagnetic oscillators is well known in the field of optics [29] and have been referred as 'chirping'. In the present paper we are interested, however, in squeezing effects in mechanical oscillators. Squeezing phenomena of mechanical oscillators have been analyzed and observed in the quantum dynamics of single trapped ions [30].

In the present work we would like to study, especially, squeezing phenomena which can occur in *mechanical oscillators* which are used as certain components in interferometers for the detection of gravitational waves on earth. A most promising source of gravitational waves, for their detection, might be given by astronomical binary systems [31] and the amount of gravitational energy emitted by such sources increases in time near their coalescence. A binary system gradually spiral inwards because of the emission of gravitational waves and the resulting waveforms increases in amplitude and frequency ('chirping effect'). It has been shown previously [5] that if energy is supplied to an oscillating system it can be simulated by a variable mass. We find, therefore, that the chirping of gravitational waves and their increased amplitudes [31] affecting the interferometric mechanical oscillators [32] can be related to the general problem of harmonic oscillator with time-dependent frequency and effective-mass. While squeezing phenomena related to the use of squeezed electromagnetic field in interferometers [33] have been extensively studied (see [34-37] which include long lists of references) the *quantum chirping effects in mechanical oscillators* used in the interferometers have been ignored. We will study in the present paper such effects.

The present paper is arranged as follows:
In Section 2 we present a general theory of harmonic oscillator with time dependent frequency and effective-mass. In Section 3 we analyze the squeezing effects produced in mechanical oscillators due to time dependent frequency ('chirping') and/or time dependent effective-mass (i.e, slow change in the harmonic oscillator amplitude due to damping or energy pumping ). In Section 4 we analyze the implications of the general theory to the mechanical oscillators used as certain components in the interferometes for detecting gravitational waves. In Section 5 we summarize our results and conclusions.

**2. The evolution operator of harmonic oscillator with time-dependent frequency and effective-mass**

We consider the following Hamiltonian of an harmonic oscillator

$$\hat{H}(t) = \frac{\hat{p}^2}{2M_{eff}(t)} + \frac{1}{2}M_{eff}(t)\omega(t)^2 \hat{q}^2 \quad , \tag{1}$$

where $M_{eff}(t)$ and $\omega(t)$ are the effective-mass and frequency, respectively, and they are time dependent. This Hamiltonian has been investigated in many articles [6-28] and we use here the analysis given in [8]. We rewrite the Hamiltonian (1) as

$$\hat{H}(t) = a_1(t)\hat{J}_+ + a_2(t)\hat{J}_0 + a_3(t)\hat{J}_- \quad , \tag{2}$$

where

$$\hat{J}_+ = \frac{1}{2\hbar}\hat{q}^2 \quad ; \quad \hat{J}_- = \frac{1}{2\hbar}\hat{p}^2 \quad ; \quad \hat{J}_0 = \frac{i}{4\hbar}(\hat{p}\hat{q} + \hat{q}\hat{p}) \quad , \tag{3}$$

and

$$a_1(t) = \hbar M_{eff}(t)\omega(t)^2 \quad ; \quad a_2(t) = 0 \quad ; \quad a_3(t) = \frac{\hbar}{M_{eff}(t)} \quad . \tag{4}$$

Here $\hat{J}_+, \hat{J}_0$ and $\hat{J}_-$ form the SU(2) Lie algebra satisfying the commutation-relations (CR):

$$\left[\hat{J}_+, \hat{J}_-\right] = 2\hat{J}_0 \quad ; \quad \left[\hat{J}_0, \hat{J}_\pm\right] = \pm\hat{J}_\pm \qquad . \tag{5}$$

The Schrodinger equation corresponding to this Hamiltonian is

$$\hat{H}(t)|\phi(t)\rangle = i\hbar \frac{\partial}{\partial t}|\phi(t)\rangle \qquad . \tag{6}$$

The evolution operator $\hat{U}(t,0)$ is given by:

$$|\phi(t)\rangle = \hat{U}(t,0)|\phi(0)\rangle \qquad , \tag{7}$$

where $|\phi(0)\rangle$ is the wavefunction at time $t=0$. Insertion of (7) into (6) gives the evolution equation

$$\hat{H}(t)\hat{U}(t,0) = i\hbar \frac{\partial}{\partial t}\hat{U}(t,0) \quad ; \quad \hat{U}(0,0) = 1 \quad . \tag{8}$$

Since $\hat{J}_+, \hat{J}_0$ and $\hat{J}_-$ form a closed Lie algebra of $SU(2)$, the evolution operator can be expressed in the following form:

$$\hat{U}(t,0) = \exp\left(c_1(t)\hat{J}_+\right)\exp\left(c_2(t)\hat{J}_0\right)\exp\left(c_3(t)\hat{J}_-\right) \quad , \tag{9}$$

where differential equations for $c_i(t)$ are obtained by direct differentiation of this operator with respect to time and by substituting the result in (8). By certain reordering of the operators on the right side of (8) and comparing them with the left side of this equation one gets three differential equations [8]:

$$\dot{c}_3 = a'_3 \exp(c_2) \quad ; \quad \dot{c}_2 = -2a'_3 c_1 \quad ; \quad \dot{c}_1 = a'_1 - a'_3 c_1^2 \quad , \tag{10}$$

with the initial conditions

$$c_1(0) = 0 \quad ; \quad c_2(0) = 0 \quad ; \quad c_3(0) = 0 \quad , \tag{11}$$

and where the $a'_j$ are defined by

$$a'_j = a_j / i\hbar \quad . \tag{12}$$

The solutions for (10) can be presented as

$$c_1(t) = iM_{eff}(t)\frac{1}{u(t)}\frac{\partial}{\partial t}u(t) \quad ; \quad c_2(t) = -\ln\frac{u(t)^2}{u(0)^2} \quad ;$$

$$c_3(t) = -iu^2(0)\int_0^t \frac{dt'}{M_{eff}(t')u^2(t')} \tag{13}$$

in which $u(t)$ satisfies the following auxiliary differential equation:

$$\ddot{u} + \gamma_0(t)\dot{u} + \omega(t)^2 u = 0 \quad , \tag{14}$$

and

$$\gamma_0(t) = \frac{\partial}{\partial t}\left[\ln(M_{eff}(t))\right] = \frac{1}{M_{eff}(t)}\frac{\partial}{\partial t}M_{eff}(t) \quad . \tag{15}$$

The above equations are similar to those derived previously (see Eqs. (4.5)-(4.8) in [8]). The present analysis is, however, more general as we assume that both the effective-mass and frequency are time dependent while in [8] the frequency was assumed to be constant. The auxiliary equation (14) for the present analysis is quite complicated and does not have a general solution. We refer to the literature [11-14] for the solution of this equation for special cases. For our purpose we shall simplify, however, our analysis in Section 4 by the use of the adiabatic approximation.

### 3. Coherence and squeezing properties of the wavefunction of harmonic oscillator with time-dependent frequency and effective-mass

The following analysis for describing the coherence and squeezing properties of harmonic oscillator with time-dependent frequency and effective-mass is based on the use of the $SU(2)$ algebra. The solutions for the time-dependent parameters $c_1(t), c_2(t)$ and $c_3(t)$ of (13) for a special case, under the adiabatic approximation, will be given in Section 4. A similar analysis to the present one is given in [20,21] by the use of $SU(1,1)$ group CR but it does not include explicit evaluation of the parameters $c_1(t), c_2(t)$ and $c_3(t)$.

Using the evolution operator (9) we can study the evolution of the wavefunction of harmonic oscillator with time-dependent frequency and effective-mass. We use the definitions:

$$\hat{q} = \sqrt{\frac{\hbar}{2\omega M}}\left(\hat{a} + \hat{a}^\dagger\right) \quad ; \quad \hat{p} = i\sqrt{\frac{\hbar\omega M}{2}}\left(\hat{a} - \hat{a}^\dagger\right) \quad , \tag{16}$$

where $\hat{a}$ and $\hat{a}^\dagger$ are the annihilation and the creation operators, respectively. Suppose we start with a coherent state at time $t = 0$:

$$|\phi(0)\rangle = |\alpha\rangle \quad , \tag{17}$$

where $|\alpha\rangle$ is the eigenstate of the annihilation operator and $\omega_0$ is the frequency at time $t = 0$. $M$ is the ordinary mass which is equivalent to the effective mass at time $t = 0$, i.e.,

$$M_{eff}(0) = M \quad . \tag{18}$$

We can define a new operator $\hat{A}$ as

$$\hat{A} = \hat{U}(t,0)\hat{a}\hat{U}^{\dagger}(t,0) \quad . \tag{19}$$

It is easy to see that the wavefunction at time $t$ is a coherent state with respect to the new operator

$$\hat{A}|\phi(t)\rangle = \alpha|\phi(t)\rangle \quad . \tag{20}$$

Using (9) and (19) it can be shown [8,20,21] that the original operator $\hat{a}$ is related to the new operator $\hat{A}$ by Bogoliubov transformation

$$\hat{A} = \eta_1 \hat{a} - \eta_2 \hat{a}^{\dagger} \quad , \tag{21}$$

with

$$|\eta_1|^2 - |\eta_2|^2 = 1 \quad . \tag{22}$$

$\eta_1$ and $\eta_2$ have been calculated [8] obtaining:

$$\eta_1 = \frac{1}{2}\exp(-\frac{c_2}{2})\left[1 + c_1 c_3 + \exp(c_2) - \frac{c_1}{M\omega_0} - M\omega_0 c_3\right] \quad , \tag{23a}$$

$$\eta_2 = \frac{1}{2}\exp(-\frac{c_2}{2})\left[1 - c_1 c_3 - \exp(c_2) + \frac{c_1}{M\omega_0} - M\omega_0 c_3\right] \quad . \tag{23b}$$

It is easy to verify that (22) is satisfied. More detailed analyses for time-dependent harmonic oscillator can be found in the literature [1-28]. For the present purpose the most interesting result is given by the uncertainty relation which is described as follows.

We define

$$\hat{x} = \frac{1}{\sqrt{2}}\left(\hat{a} + \hat{a}^{\dagger}\right) \quad , \tag{24a}$$

$$\hat{p} = \frac{1}{i\sqrt{2}}\left(\hat{a} - \hat{a}^{\dagger}\right) \quad . \tag{24b}$$

Then by straightforward but somewhat lengthy calculations one finds [8] :

$$\Delta x^2 = \langle \hat{x}^2 \rangle - \langle \hat{x} \rangle^2 = \frac{1}{2}\exp(-c_2)\left(1 - M^2 \omega_0^2 c_3^2\right) \quad , \tag{25}$$

$$\Delta p^2 = \langle \hat{p}^2 \rangle - \langle \hat{p} \rangle^2 = \frac{1}{2}\exp(-c_2)\left\{\left[c_1 c_3 + \exp(c_2)\right]^2 - \frac{c_1}{M^2 \omega_0^2}\right\} \quad . \tag{26}$$

We find that the most important terms in (25-26) are given by

$$\Delta x \propto \exp\left(-\frac{c_2}{2}\right) \quad , \tag{27}$$

$$\Delta p \propto \exp\left(\frac{c_2}{2}\right) \quad . \tag{28}$$

So, squeezing in the fluctuations of one operator in the expense of an increase in the fluctuations of the other operator is obtained. The approximate relations (27-28) have been derived also by other authors (see e.g., [21], Eqs. (92-93)). In the present calculations we have not taken into account driving forces since the present model is applied particularly in Section 4 to tidal gravitational waves operating on interferometric detectors where the use of the Hamiltonian (5) will be exploited. Also as is apparent from previous works [20-22] the effect of the driving force is to increase the signal but it does not change the uncertainty relations given by the approximate relations (27-28). The critical point in estimating the magnitude of the squeezing phenomena is the explicit calculation of the parameters $c_1(t), c_2(t)$ and $c_3(t)$. As shown in the previous section this can be done only by solving the auxiliary differential equation (14) which has no general solution but can be solved either for specific cases or by assuming certain approximations.

**4. Gravitational 'tidal' forces operating on 'chirped' mechanical oscillators**

In the following discussion we review shortly the use of inteferometers for measuring gravitational waves [31]. Then we show that 'chirping' effects including increased amplitude, for gravitational waves on earth operating on certain components of the interferometers, can be related to previous theories of time-dependent effective-mass and frequency of harmonic oscillators.

Gravitational waves are propagating fluctuations of gravitational fields, that is ,"ripples " in space time which travel with the speed of light. Every body in the path of such a wave feels a 'tidal' gravitational force that acts perpendicular to the waves direction of propagation; these forces change the distance between points, and the size of the changes is proportional to the distance between these points thus gravitational waves can be detected by devices which measure the induced length changes. A promising form of gravitational wave detector uses laser beams to measure the distance between two well- separated masses. Such devices are basicly kilometer sized laser interferometers consisting of three masses placed in L-shaped configuration. The laser beams are reflected back and forth between the mirrors attached to the masses, where the mirrors lying several kilometers away from each other. A gravitational wave passing through this interferometer will cause the length of the arms to oscillate with time. For a polarized gravitational when one arm contracts the other expands and this pattern alternates. The result is that the interference pattern of the two laser beams changes with time. It is expected that laser interferometric detectors are the ones that will provide us with the first direct detection of gravitational waves on earth.

One way of understanding the effects of gravitational waves operating on interferometers is to describe them as *tidal forces* operating on test masses (corresponding to the masses on which the interferometric mirrors are placed). Imagine that we have placed two freely-falling test particles of mass $m$ along the $x$ axis, at $x = \pm L$ (where $L$ is small compared to the wavelength of the gravitational wave). If a gravitational wave is incident along the $z$ axis in the + polarization the distance change between the masses is equal to their distance multipled by $h_{11}/2$ [31,32,38,39]. The stronger the gravitational wave, the greater the relative length change, or *strain* that will occur between two points in spacetime. The amplitude of the gravitational wave $h_{11}$ is defined as twice the strain. The *classical* equation of motion for the test-mass (neglecting any damping effects due to external noise e.g., seismic noise) is given by

$$m\frac{d^2 \Delta x}{dt^2} = F_{gw} = \frac{1}{2} mL \frac{\partial^2 h_{11}}{\partial t^2} \quad . \tag{29}$$

We find that the change $\Delta x$, due to the gravitational wave, in the location of the mass $m$ is given by $h_{11}$ multiplied by certain proportionality constants including the interferometer arm length. We have simplified the analysis by assuming a certain polarization for the gravitational wave but the analysis can be generalized by taking into account other gravitational waves' polarizations and more general conditions [31,32,38-40] .

Quite often it is claimed that the time-dependence of the strain is harmonic so that

$$h_{11} = h_{11}^{(0)} \cos(\omega t + \phi) \quad ; \tag{30a}$$

$$\Delta x = \frac{1}{2} L h_{11}^{(0)} \cos(\omega t + \phi) \quad ; \tag{30b}$$

$$\frac{\partial^2 h_{11}}{\partial t^2} = -\omega^2 h_{11}^{(0)} \cos(\omega t + \phi) \quad , \tag{30c}$$

where $h_{11}^{(0)}, \omega$ and $\phi$ are constants. For an interferometer with arm length of 4 km we are looking for maximal changes $\Delta x_0$ in the arm length of the order of [41]

$$\Delta x_0 = h_{11} L \approx 10^{-22} \times 4 km = 4 \times 10^{-17} cm \quad ! \tag{31}$$

This small number explains why all detection efforts with ultra-sensitive detectors till today were not successful.

Physical experiments with single macroscopic objects became so accurate that quantum mechanical uncertainty fluctuations should be taken into account. By simple considerations (see e.g . [42]) the minimum quantum mechanical error in the particle position measurement during time of measurement $\tau$ is given by the 'standard quantum limit' (SQL):

$$\Delta x_{SQL} = \sqrt{\frac{\hbar \tau}{2m}} \quad . \tag{32}$$

Masses of order some kg are used in gravitational waves detectors and much effort is spent to eliminate the coupling of the system with 'seismic' noise. The measurement is usually expected to be made during a time which is smaller than the time period of the gravitational wave so that the effect of gravitational wave does not alternate between the two arms of the interferometer. Taking this into account another form for the SQL has been given [42]:

$$\Delta x_{SQL} = \sqrt{\frac{\hbar}{2m\omega}} \quad , \tag{33}$$

where $\omega$ is the frequency of the gravitational wave. Putting some numbers in a typical example we have: $\hbar \approx 10^{-27} erg.\sec.$; $m = 10^4 gm$; $\omega = 100 \, cycles / \sec$, then the SQL is approximately: $\Delta x_{SQL} \approx 2 \times 10^{-17} cm$, which is of the same order of magnitude as the signal of (31). Of course the time of measurement can be much smaller than the time period of the gravitational wave, improving the SQL, but usually in the experiments the signal is increased by multi-reflections so that the SQL cannot be ignored. From this rough calculation one can conclude that quantum mechanical fluctuations are important for gravitational waves detectors. In the early works on detection of gravitational waves [33,42] the SQL was considered as a certain quantum limit for observing small position displacements but in later works it has been shown that by using squeezed states of radiation (see e.g. [34-35] and the lists of References included in these articles) or by other methods (see e.g., [37] and the list of References in this article) that it is possible to observe position displacements which are smaller than the SQL.

In the present work we would like to analyze another effect produced by gravitational waves on earth which has been ignored in previous works. Let us assume that the gravitational wave is emitted from an astronomical binary-system where the gravitational wave and its 'tidal' forces on earth satisfy the harmonic-oscillator equations *as long as the binary-system is far from its coalescence*. As the binary system spiral inwards there are deviations from the harmonic oscillator equations of (30): a) The strain amplitude is increasing as a function of time. In the adiabatic approximation the variation of the time dependent amplitude is long compared to the characteristic time period of the system. Under this approximation $h_{11}^{(0)}$ of (30a) can be exchanged into

$$h_{11}^{(0)}(t) = f(t)h_{11}^{(0)} \quad ; \quad f(0) = 1 \quad , \tag{34}$$

where $f(t)$ represents a slowly growing amplitude of time with initial function $f(0) = 1$, i.e., $h_{11}$ increases as the binary system spirals inwards [31,32]. b) The frequency of $h_{11}$ increases when the binary system spiral inward (chirping effect [31]). Under the adiabatic approximation $\cos(\omega t + \phi)$ of (30a) can be exchanged into

$$\cos(\omega(t)t + \phi_0) \quad ; \quad \omega(0) = \omega_0 \quad , \tag{35}$$

where $\omega(t)$ is a slowly varying function of time. These approximations describe classical equations of motion which start as harmonic functions with frequency $\omega_0$ and initial phase $\phi_0$. As the binary system spiral inwards both the amplitude and frequency are increasing as function of time.

Using the changes (34-35), equation (30a) is transformed under the adiabatic approximation into

$$h_{11}(t) = f(t)h_{11}^{(0)} \cos(\omega(t)t + \phi_0) \quad . \tag{36}$$

We define the variable q as

$$q = \Delta x = \frac{1}{2} L f(t) h_{11}^{(0)} \cos(\omega(t)t + \phi_0) \quad . \tag{37}$$

Within the adiabatic approximation the displacement $q = \Delta x$ follows the changes in the amplitude and frequency of the gravitational wave and its time derivative is given by

$$\dot{q} = -\frac{1}{2} L f(t) h_{11}^{(0)} \omega(t) \sin(\omega(t)t + \phi_0) \quad . \tag{38}$$

We have neglected in (38) the derivative of $f(t)$ and $\omega(t)$ according to time but have taken into account their slow change with time. (In the adiabatic approximation the changes of these variables are quite small during a time period of oscillation and their changes over many periods is taken into account).

In order to quantize the equations of motion we introduce the definition

$$M_{eff} = \frac{M}{f(t)} \quad , \tag{39}$$

where M is the ordinary mass and $M_{eff}$ is defined as the effective-mass. Then the equation of motion (38) becomes

$$\dot{q} = \frac{p}{M_{eff}} \quad ; \quad p = -\frac{1}{2} L f(t) M \, h_{11}^{(0)} \omega(t) \sin(\omega(t)t + \phi_0) \quad , \tag{40}$$

where $p$ is the linear momentum. The equation of motion for p, under the adiabatic approximation, becomes

$$\dot{p} = -\frac{1}{2} L f(t) M \, h_{11}^{(0)} \omega(t)^2 \cos(\omega(t)t + \phi_0) = -M_{eff}(t) \omega(t)^2 q \quad . \tag{41}$$

The quantization of the present one dimensional harmonic oscillator is given by the Hamiltonian operator

$$\hat{H}(t) = \frac{\hat{p}^2}{2 M_{eff}(t)} + \frac{1}{2} M_{eff}(t) \omega(t)^2 \hat{q}^2 \quad , \tag{42}$$

which is equivalent to the Hamiltonian (1). One should, however, take into account that in our analysis the effective mass $M_{eff}(t)$ is smaller than M as it simulates energy pumping [5] (increasing amplitude) while in most treatments of the Hamiltonian (1) it simulates energy damping (decreasing amplitude).

The variables $\hat{q}$ and $\hat{p}$ have become operators satisfying the CR

$$[\hat{q}, \hat{p}] = i\hbar \quad , \tag{43}$$

and the canonical equations of motion are

$$\dot{\hat{q}} = \frac{1}{i\hbar}\left[\hat{q}, \hat{H}\right] = \frac{p}{M_{eff}} \quad ; \quad \dot{\hat{p}} = \frac{1}{i\hbar}\left[\hat{p}, \hat{H}\right] = -M_{eff}(t)\omega(t)^2 q \quad . \tag{44}$$

The quantized equations of motion (44) correspond to the classical equations of motion (40-41). While $\hat{q}$ and $\hat{p}$ are considered as operators, $M_{eff}(t)$ and $\omega(t)$ are taken into account as classical variables.

Since the quantization of the present system is similar to the general analyses of time-dependent harmonic oscillators [1-28] we expect that there will be quantum squeezed noise effects in the present system which will be analogous to those analyzed previously [8-28] in other systems. We will estimate the magnitude of squeezing effects obtained by the time-dependent gravitational waves 'tidal' forces, by using the following analysis.

For calculating the time-dependent parameters $c_1(t), c_2(t)$ and $c_3(t)$ of (13) we need first to find the solution for the auxiliary equation given by (14-15) which as we show simulates the classical second order differential equation for q. Using (40) we get

$$\ddot{q} = \frac{\dot{p}}{M_{eff}} - p\frac{1}{M_{eff}^2}\frac{dM_{eff}}{dt} \qquad M_{eff} = \frac{M}{f(t)} \tag{45}$$

Substituting in (45), $\dot{p}$ from (41), $p$ from (40) and $\gamma_0$ from (15) then we get

$$\ddot{q} + \gamma_0 \dot{q} + \omega(t)^2 q = 0 \qquad . \tag{46}$$

so that the differential equation for the classical q variable, within the adiabatic approximation, is equivalent to the differential equation for u.

In performing the first order time derivative of $q$, or correspondingly $u$, we have neglected the time derivative of $f(t)$. In performing the second order time derivative of $q$, or correspondingly $u$, we have taken into account the first order time derivative of $f(t)$ multiplied by $\omega(t)$. This term is considered as a correction term proportional to $\gamma_0$ and $\dot{q}$. This approximation is obtained within the adiabatic approximation and is consistent with the above quantization procedure.

We use the relations

$$u(t) = u_0 f(t)\cos(\omega(t) + \phi_0) \quad ; \quad f(t) = \frac{M}{M_{eff}} \quad , \tag{47}$$

where $u(t)$ is the solution of (14) within the adiabatic approximation and where $f(0) = 1$. Substituting (47) into (13) we get

$$c_1(t) = iM_{eff}(t)\frac{\partial}{\partial t}\{\ln[f(t)\cos(\omega(t)+\phi_0)]\} \quad , \tag{48a}$$

$$c_2(t) = -\ln\{[f(t)\cos(\omega(t)+\phi_0)]^2/\cos^2\phi_0\} \quad , \tag{48b}$$

$$c_3(t) = -i\int_0^t \frac{\cos^2(\phi_0)dt'}{M_{eff}(t')[f(t')\cos(\omega(t')+\phi_0)]^2} \quad . \tag{48c}$$

Using cgs units we should notice that in Eqs. (13) or correspondingly (48), $c_1(t)$ has the dimension $\frac{gram}{sec}$, $c_3(t)$ has the dimension $\frac{sec}{gram}$, while $c_2(t)$ is dimensionless. Vice versa, in the definitions (3) $\hat{J}_+$ has the dimension $\frac{sec}{gram}$, $\hat{J}_-$ has the dimension $\frac{gram}{sec}$ while $\hat{J}_0$ is dimensionless .

In order to estimate the effects of $c_1(t)$, $c_2(t)$ and $c_3(t)$ on the time evolution of the harmonic-oscillator we consider first the effect of $c_1(t)$. We find by performing the derivative in (48a) that this derivative includes in the denominator the function $f(t)$ and additional function $f(t)^{-1}$ appears from the definition of $M_{eff}$ of (45) .The other terms are relatively small and oscillating with time. Since we are considering harmonic oscillator in which $f(t) \gg 1$ this term tends to zero with increased time near the coalescence of the binary system.

For estimating the value of $c_3(t)$ we can average the rapidly oscillating terms in the integrand over a time period obtaining

$$c_3(t) = -i\int_0^t \frac{1}{M_{eff}(t')f(t')^2} \quad . \tag{49}$$

The relative contribution of terms proportional to $c_3(t)$ to the uncertainties of (25-26) is relatively small since the integrand of $c_3(t)$ becomes proportional to $\frac{1}{f(t')}$ (taking into account (39)) and as we are treating the limits with very large values of values of $f(t)$. It is ,however, not completely negligible.

By averaging the the rapidly oscillating terms in (48b) we get

$$c_2(t) \approx -2\ln(f(t)) \qquad . \qquad (50)$$

Within the present approximations we find according to (27-28) that the quantum uncertainty of one quadrature is increasing proportional to $f$ while the quantum uncertainty in the other quadrature is decreasing proportional to $1/f$. This result is not surprising as it is consistent with all others works on the quantum noise in time-dependent harmonic-oscillators. It is quite easy to find that all the terms of (25-26) which included the parameters $c_1(t)$ and $c_3(t)$ and were considered to be relatively small are *positive so that their effects will be to increase the quantum* noise in both quadratures. The physical conclusion which comes out from the above analysis is that *if we cannot separate experimentally the two quadratures of the mechanical* oscillator detector then the quantum noise will increase proportional to the increase of the gravitational wave intensity as it approaches near the coalescence point of the binary system. On the other hand it might be interesting to study the possibility of separating the two quadratures of the mechanical oscillators in such interference experiments.

## 5. Sumary, discussion and conclusion

The theory of harmonic oscillators with time-dependent frequency and effective-mass have been developed in previous studies [1-28] . Such quantum theories have been used for explaining damping and pumping effects in non-conservative systems. It has been shown also that such systems will produce squeezing phenomena ,i.e, the quantum fluctuations in one quadrature will be reduced on the expense of increasing the quantum noise in the other quadrature. A general theory for such effects has been given in the present paper on Sections 2-3.

The general theory for time-dependent harmonic oscillator has been applied in the present research for studying certain quantum effects in the interferometers for detecting gravitational waves. 'Tidal' gravitational waves forces lead to displacements of the masses on which the interferometric mirrors are placed causing the length of the interferometric arms to oscillate with time. For a polarized gravitational wave when one arm contracts the other expands and this pattern alternates. The result is that the interference pattern of the laser beams changes with time and it is hoped to be able to detect the gravitational wave by detecting such interference changes. The magnitude of the arms length changes is extremely small so that physical experiments with single macroscopic objects became so accurate that quantum mechanical uncertainty fluctuations should be taken into account. Quite often it is claimed that the time-dependence of the changes in the arm length is harmonic. However, for the detection of gravitational wave on earth by interferometers one looks for strong gravitational waves sources. Such sources can be obtained by binary astronomical sources near their merging point. As the binary system approaches its point of coalescence the gravitational wave intensity and frequency are increasing and this can lead to strong deviations from the simple harmonic oscillations for the interferometric masses. It is shown in the present article that under such conditions a better description for the mirrors displacements will be given by an harmonic oscillator with time-dependent frequency and amplitude.

The time-dependent mass (the 'effective-mass') in which the mass is increased with time usually describes damping phenomena . In the present theoretical model the effective-mass is decreasing with time describing pumping phenomena in which the oscillator amplitude is increasing with time. The present theoretical model of the harmonic oscillator is developed within the 'adiabatic' approximation by which the displacements of the mirrors follow the changes in the amplitude and frequency of the gravitational wave. Under such condition it is shown that the harmonic oscillations of the mirrors can be described by harmonic-oscillators with time dependent frequency and effective-mass. The quantization of such systems are analyzed. It is found that the increase of the gravitational wave intensity within the adiabatic approximation leads to squeezing phenomena where the noise in one quadrature is increased and in the other is reduced.

**Acknowledgement**
The author would like to thank Amos Ori for interesting discussions